\documentclass{PoS}
\usepackage{url}

\title{Polarimetry with POLAR}

\ShortTitle{Performance of the POLAR detector}

\author{\speaker{Merlin Kole on behalf of the POLAR collaboration}\thanks{As presented: on isdc.unige.ch/polar/}\\
       DPNC, University of Geneva\\
        E-mail: \email{merlin.kole@unige.ch}}

\abstract{The year 2017 marked the 50th anniversary of the detection of the first Gamma-Ray Burst. The 5 decades following the discovery have seen a large interest in these transient phenomena accompanied by both dedicated space and ground-borne detectors to study the emission over a wide energy range. However, many questions on the progenitors of the bursts, the emission processes involved and the emission environments still remain. Some of the most important remaining questions can be solved by probing one of the remaining unstudied parameters of the high energy component of the emission, the polarization. POLAR is a space-borne detector designed specifically to measure the polarization of the prompt emission of GRBs in the 50-500 keV energy range. For this purpose the instrument uses a segmented scintillator array to measure the potential asymmetry in the azimuthal Compton scattering angle induced by the polarized component of the photon flux. The POLAR instrument was launched as part of the second Chinese spacelab, the Tiangong-2 on the 15th of September 2016 from the Jiuquan Satellite Launching Centre in Inner-Mongolia, China. Due to its relatively small pixel size POLAR is able to measure the photon interaction locations, and therefore the scattering angles, with a high precision resulting in a relatively high sensitivity to polarization. The instrument furthermore has a large effective area and a field of view of approximately 1/2 the sky, making POLAR optimal for detecting transient events and making it one of the most sensitive detectors currently in orbit in its energy range. In the first half year of operation POLAR detected a total of 55 Gamma-Ray Bursts about 10 of which were bright enough to allow for detailed polarization studies, thereby forming the start of the first Gamma-Ray Burst polarization catalog. In this paper a brief overview of the previous GRB polarization studies will be presented followed by an overview of the POLAR detector along with the first result of the in-flight performance. The detected Gamma-Ray bursts will be presented and finally prospects for polarization measurements of these events will be discussed.  }

\FullConference{XII Multifrequency Behaviour of High Energy Cosmic Sources Workshop\\
		12-17 June, 2017\\
		Palermo, Italy}

\begin{document}

\section{Introduction}

While the intensity, the timing and the spectral features of the $\gamma$-ray emission have been studied in great detail for a large variety of astrophysical sources, the polarization degree and angle remain largely unprobed. As the polarization parameters are however highly dependent on the environment in which the $\gamma$-ray emission takes place, the information held by these parameters can be used to answer a wide range of questions for different sources such as pulsars, active galactic nuclei and Gamma-Ray Bursts (GRBs). 

Thanks to the large scientific potential of polarimetry, indicated by the large number of contributions on this topic at previous editions of this workshop, and the improvements in technology, a number of new missions dedicated to polarimetry have appeared in recent years. Some examples of large scale planned missions are IXPE \cite{IXPE} and eXTP \cite{eXTP} while several smaller scale missions, such as the GAP detector \cite{GAPGRB}\cite{GAPGRB2} and the balloon-borne COSI \cite{COSI} and PoGOLite detectors \cite{kole} have already performed polarization measurements in recent years. In addition some attempts have been made with instruments without a dedicated polarimeter to perform polarization measurements of astrophysical sources in the X/$\gamma$ ray energy range. Examples of this are measurements of the Crab pulsar by two different detectors on the INTEGRAL mission  \cite{Max} \cite{IBIS}. 

The POLAR instrument \cite{Nicolas} is a new dedicated GRB polarimeter launched as part of the Chinese Tiangong 2 (TG-2) Space Lab on September 15th 2016. The goal of the mission is to produce a large catalog of detailed polarization measurements of GRBs in the $50-500\,\mathrm{keV}$ energy range. In the following section the scientific goals of POLAR are briefly discussed along with the history of GRB polarization measurements at high energies. This is followed by a description of the instrument and its performance in-orbit. Finally an overview of the measurements performed by the instrument during its first months of operation will be given and the prospects for polarization measurements will be provided.

\section{Gamma-Ray Bursts}

Gamma-Ray Bursts are the brightest events in the universe since the big-bang consisting of a short flash of X/$\gamma$-rays, called the prompt-phase, followed by a longer lasting afterglow at lower energies. The progenitors of these bright transient events are thought to be the death of massive stars in the case of long GRBs, which typically last from seconds to minutes, and the merger of two compact objects in the case of short GRBs which typically last less than 2 seconds.  The association of short GRBs with the merger of two compact objects was confirmed in the summer of 2017 with the first combined detection of gravitational waves by LIGO and VIRGO and the electromagnetic component by both Fermi-GBM and INTEGRAL SPI-ACS \cite{LIGO}. According to most emission models the high energy component observed from both long and short GRBs originates from two jets. However, the exact processes involved in the emission and details on the emission environment such as the importance of a magnetic field in the emission are still open questions.

Although difficult to differentiate using spectral measurements, the different proposed emission mechanisms predict different characteristics in the polarization parameters. Although for details on the polarization parameters predicted by different models the reader is referred to for example \cite{Toma}, a rough example of the use of polarization for gaining an understanding on the emission mechanisms can be provided here. Regarding the polarization degree, models based on emission through synchrotron radiation in a strong highly ordered magnetic field predict an average polarization degree of 40\% within a sample of GRBs \cite{Luti}. Photospheric emission models based on inverse-Compton scattering on the other hand predict only high polarization degrees when observing the burst at a large angle, the average polarization degree is therefore expected to be close to $0\%$ while polarization degrees higher than $30\%$ will be rare \cite{ChrisL}. The same strong negative correlation between the observing angle the and polarization degree is predicted by the also inverse-Compton scattering dominated Cannon Ball models \cite{Dar}, however here the maximum achievable polarization degree reaches close to $100\%$. It is clear that in order to distinguish between such models a polarization measurement of a single GRB does not suffice. rather the distribution of the polarization degrees of a large sample of GRBs needs to be produced in order to start rejecting emission models.

To date a limited number of measurements has been performed of the polarization of the prompt emission of GRBs. A list of those published is presented in table \ref{tab:title}, together with the measured polarization degree and the error on the measurement. From this list it is clear that the typical error is relatively large while for a number of measurements separate analyses of the data resulted in different results. This is a consequence of the fact that most of the listed instruments were not designed as polarimeters and were therefore not calibrated on ground. This makes analysis challenging as instrumental systematic have the potential to dominate the measurements, as is for example discussed in \cite{SPI1} with respect to their measurement of 041219A with the non-dedicated SPI instrument. The list indicates that in order to be able to distinguish between different emission models one needs a large catalog of more detailed measurements provided by dedicated instruments which have undergone detailed on ground and in-orbit calibration. 

\begin {table}
\begin{center}
  \begin{tabular}{ | l | c | c | c |}
    \hline
    \textbf{GRB} & \textbf{Instr./Sat.} & \textbf{Pol. ($\%$)} & \textbf{Ref.} \\ \hline
    160530A & COSI & $<46\%$ & \cite{COSI2}  \\ \hline
    110721A & GAP/IKAROS & $84^{+16}_{-28}$ & \cite{GAPGRB2} \\ \hline
    110301A & GAP/IKAROS & $70\pm22$ & \cite{GAPGRB2}  \\ \hline
    100826A & GAP/IKAROS & $27\pm11$ & \cite{GAPGRB} \\ \hline \hline
    021206 & RHESSI & $80\pm20$ & \cite{RHESSI2}  \\ \hline
    021206 & RHESSI & $41^{+57}_{-44}$ & \cite{RHESSI1}  \\ \hline \hline

    140206A & IBIS/INTEGRAL & $\ge48$ & \cite{IBIS1}  \\ \hline
    061122 & IBIS/INTEGRAL & $\ge60$ &  \cite{IBIS3} \\ \hline
    041219A & IBIS/INTEGRAL & $\le4 /  43\pm25$ & \cite{IBIS2}  \\ \hline
    041219A & SPI/INTEGRAL & $98\pm33$ & \cite{SPI1}  \\ \hline \hline
    960924 & BATSE/CGRO & $\ge50$ & \cite{BATSE1}  \\ \hline
    930131 & BATSE/CGRO & $\ge35$ & \cite{BATSE1}  \\ \hline

  \end{tabular}
  \caption {List of published GRB polarization measurements} \label{tab:title} 
\end{center}
\end {table}

\section{POLAR}

The POLAR detector is specifically designed to measure the polarization of the promp emission of GRBs in the 50-500 keV energy range. In order to measure the polarization parameters the instrument is designed to accurately measure the azimuthal scattering angle of the incoming photons. This azimuthal scattering angle is related to the polarization angle as can be seen in the Klein-Nishina equation \cite{KNF}: 

\begin{equation} \label{KN} 
\mathrm{\frac{d\sigma}{d\Omega}=\frac{r_0^2}{2}\frac{E'^2}{E^2}\left(\frac{E'}{E}+\frac{E}{E'}-2\sin^2\theta \cos^2\phi\right).}
\end{equation}

\begin{figure}[h!]
  \centering
    \includegraphics[width=5 cm]{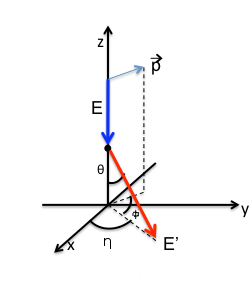}
  \caption[Schematic representation of the Compton scattering process.]
{Schematic representation of the Compton scattering process. The incoming photon (blue;
energy E, polarization $\vec{p}$) scatters off an electron with a polar angle $\theta$ from its original trajectory. $\phi$ is the angle between the polarization vector of the incoming photon and the azimuthal scattering angle $\eta$. Taken from \cite{PhD}. }
\label{polcomp}
\end{figure}

Where $\mathrm{r_0^2}$ is the classical electron radius, $\mathrm{E}$ is the initial energy of the photon, $\mathrm{E'}$ the final energy of the photon, $\theta$ is the polar scattering angle and $\phi$ the azimuthal Compton scattering angle with respect to the polarization vector as visualized in figure \ref{polcomp}. As a result of the $\cos^2\phi$ term  photons preferentially scatter perpendicular to their initial polarization vector, causing the azimuthal scattering angle to be modulated with a 180 degree period for a polarized flux. The distribution of the azimuthal scattering angles will follow a function of the form:

\[f(\phi)=A(1+\mu \cos(2(\eta-\phi)+\pi)\]

Here $\phi$ is the angle between the polarization vector of the incoming photon and the scattering angle, $\eta$ is the measured scattering angle, and $\mu$ is the amplitude of the harmonic referred to as the modulation factor, while $A$ is simply the mean of the harmonic. The polarization degree can be retrieved by dividing $\mu$ by $\mu_{100}$ which is the amplitude of the harmonic measured using the instrument for a $100\%$ polarized beam. The value of $\mu_{100}$ is an instrument dependent variable which furthermore depends on the energy of the incoming photons and the incoming angle of the flux with respect to the instrument. In the case of POLAR this value is acquired using simulations verified using  dedicated on ground calibration tests \cite{NIM} and is of the order of $30-40\%$ depending on the photon energy. A last parameter often used in polarization to indicate the statistical significance of a measurement is the Minimal Detectable Polarization (MDP) \cite{Weisskopf}. The MDP represents the minimum degree of polarization required for the flux in order to confirm the flux not to be unpolarized with $3\sigma$ confidence.

\begin{figure}[h!]
  \centering
    \includegraphics[width=8 cm]{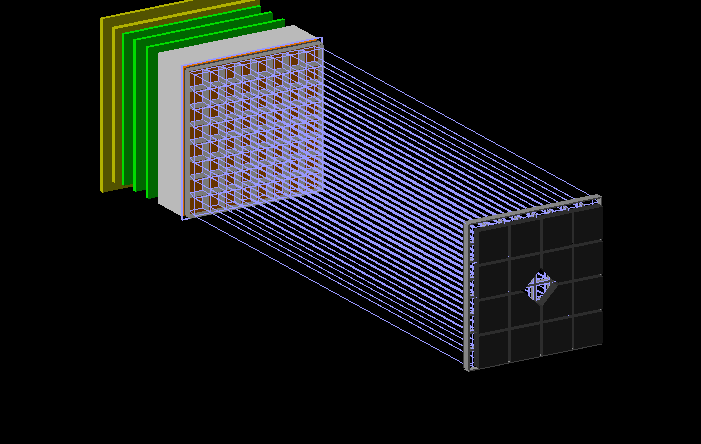}
  \caption[Schematic representation of the Compton scattering process.]
{Schematic representation of a single module of POLAR. The scintillator bars, shown in blue, together with the MAPMT (grey), the front-end electronics (green) and the rubber dampers on top (black). Taken from \cite{NIM}.}
\label{module1}
\end{figure}

\begin{figure}[h!]
  \centering
    \includegraphics[width=8 cm]{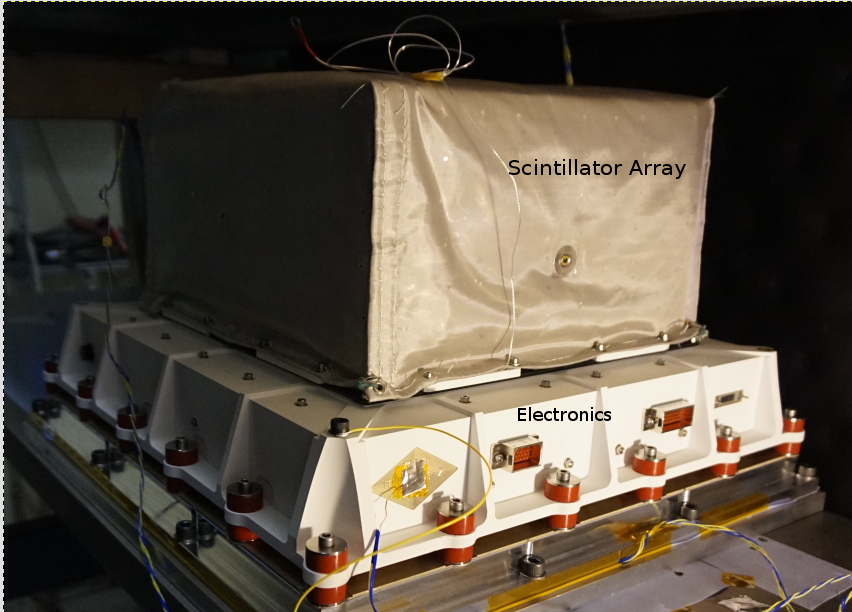}
  \caption[Schematic representation of the Compton scattering process.]
{The POLAR flight model during thermal vacuum tests. The electronics is contained in the white painted aluminum structure while the scintillator array is placed in a carbon structure covered with a multi-layer insulator. Taken from \cite{ICRC}.}
\label{FM}
\end{figure}

In order to measure the azimuthal Compton scattering angle, POLAR uses a total of 1600 plastic scintillator bars with a size of $5.8\times5.8\times176\,\mathrm{mm^3}$. The bars are used to detect both the Compton scattering interaction and the subsequent interaction of the photon which can either again be Compton scattering or photo-absorption. The scattering angle is then reconstructed by taking the position of the two bars in which an energy deposition was measured. The 1600 bars are read out in groups of 64 by 25 Multi-Anode PhotoMultiplier Tubes  (MAPMT) each with their own front-end electronics (FEEs), the combination of the FEE, MAPMT and the bars is referred to as a module. The 25 independent detector modules communicate with central FPGAs which handle the overall trigger logic and the communication with the space station. A single detector module consisting of 64 bars, a MAPMT and the FEE is shown in figure \ref{module1}. The full flight model is shown in figure \ref{FM}. A full, detailed overview of the instrument design and construction is provided in \cite{Nicolas2}.

The instrument was launched on the 15th of September 2016 from the Jiuquan Satellite Launching Centre in Inner-Mongolia, China, as part of the Tiangong 2 (TG-2) space laboratory. The trajectory of TG-2 and the position of POLAR on top of it is such that it always points to zenith, therefore POLAR continuously observes half the sky. The instrument commenced data taking successfully on the 23rd of September 2016. All 1600 channels were found to operate without problems and the noise levels observed in-orbit are largely equal to those measured during pre-launch tests \cite{Nicolas2}. The background induced trigger rates and their dependencies on the position of TG-2 with respect to the Earth were also found to largely match those predicted by simulations presented in \cite{estella}. Finally the observed rate of GRBs detected by POLAR of approximately 10 per month matched that predicted prior to launch. It was furthermore found that POLAR is one of the most sensitive instruments in its energy range with an effective area of approximately  $300\,\mathrm{cm^2}$ in the $20-500\,\mathrm{keV}$ energy range making it capable to detect very faint GRBs potentially undetected by other instruments. In April 2017 a problem occurred in the high voltage power supply making measurements impossible from this moment until the time of writing. Attempts to recover the performance of the high voltage power supply are ongoing.

\section{GRBs Measurements and Future Prospects}

\begin{figure}[h!]
  \centering
    \includegraphics[width=15 cm]{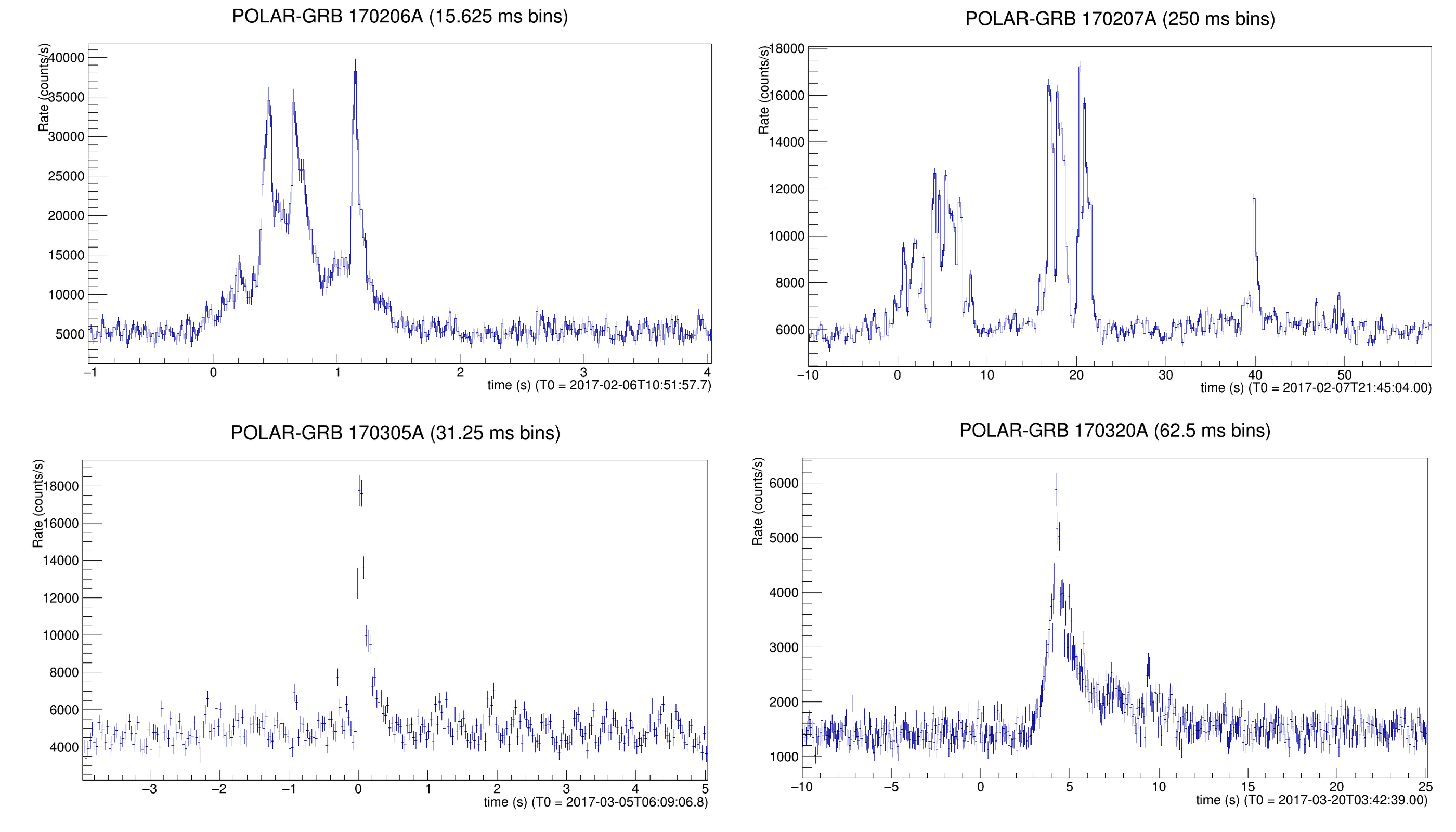}
  \caption[The light curves of 4 GRBs measured by POLAR. ]
{Examples of the light curves of 4 different GRBs measured by POLAR. All light curves measured by POLAR can be found on \url{http://www.isdc.unige.ch/polar/lc/}.}
\label{LCs}
\end{figure}

To date POLAR has a total of 55 confirmed GRB detections. Examples of several light curves measured by POLAR of both short and long GRBs are shown in figure \ref{LCs}. All the light curves can be found on \url{http://www.isdc.unige.ch/polar/lc/}. While the majority of these GRBs are faint and have a large MDP, meaning they are too weak to allow for significant polarization analysis, approximately 10 of the GRBs observed by POLAR have an MDP lower than $30\%$. Furthermore several of these GRBs have an MDP low enough to allow for potential studies of the variation of the polarization with time, allowing to study potential changes in for example the magnetic field within the jets. The brightest sample of GRBs would allow to extend the list presented in section 2 of this paper with an additional 10 GRBs all of which with equal or smaller errors as the most accurate measurements presented in that list. It should be noted here, however, that the value of $\mu_{100}$, the sensitivity to polarization, of POLAR is dependent both on the spectral shape and on the incoming angle of the GRB with respect to POLAR. The exact values of the MDP, which in turn is dependent on the value of $\mu_{100}$ therefore requires detailed studies using the spectra and location of the GRB. An overview of all the GRBs detected by POLAR is presented in \cite{Shaolin} where furthermore a list of preliminary MDP calculations is provided.

\begin{figure}[h!]
  \centering
    \includegraphics[width=10 cm]{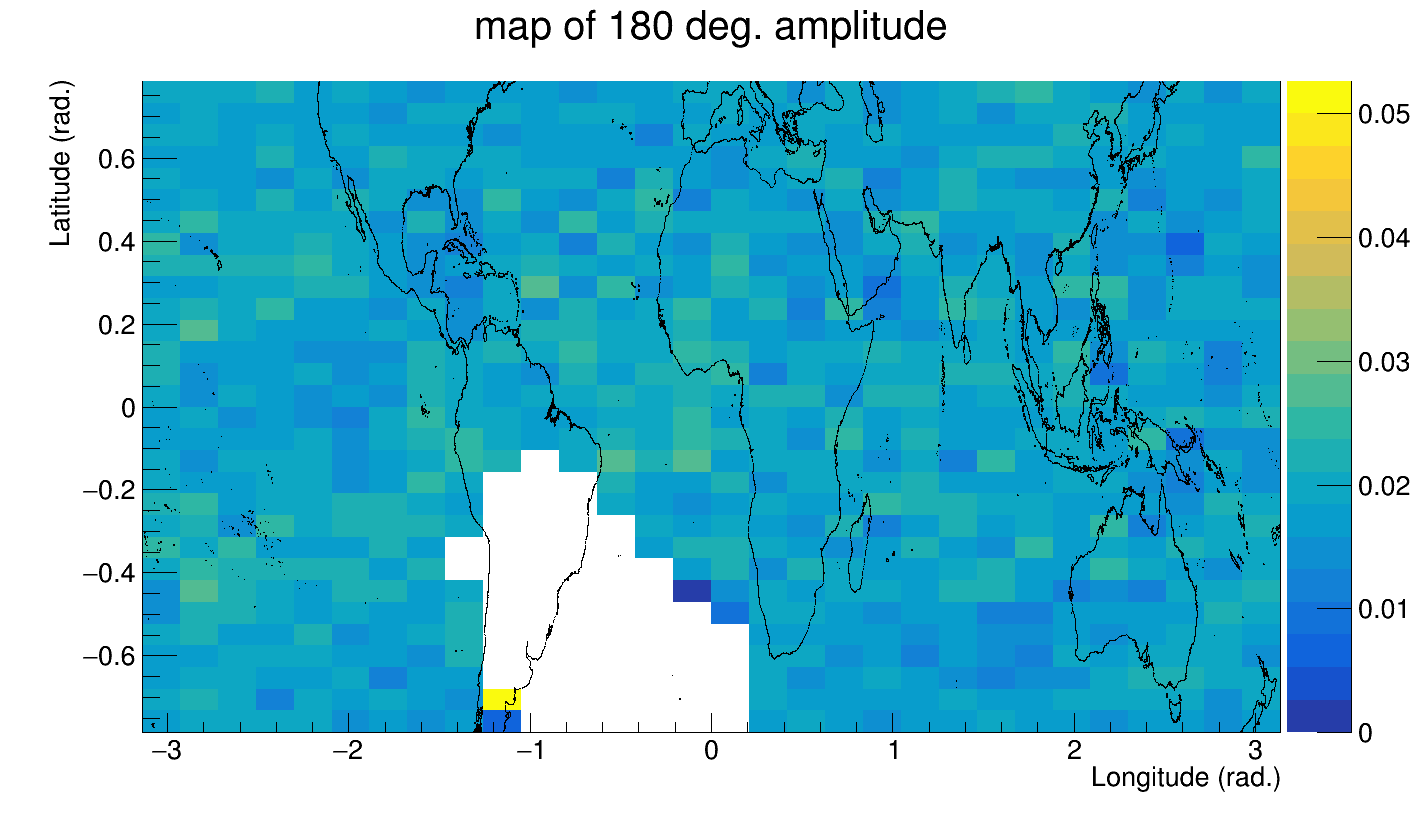}
  \caption[The background induced 180 degree modulation as a function of position.]
{The amplitude of the 180 degree modulation as measured from background data by POLAR as a function the position of the instrument with respect to Earth. The amplitude can be seen to be very stable. The empty region in the figure coincides with the South Atlantic Anomaly where the instrument does not take data.}
\label{bg_mod}
\end{figure}

The acquired statistics for the brightest GRBs allows for detailed polarization measurements, however, in order to accurately perform these measurements a detailed understanding of the instrument performance and its systematics is required. For example the potential induction of a fake polarization signal by instrument systematics needs to be studied in detail first as discussed earlier in this paper. For this purpose POLAR underwent detailed calibration tests prior to launch, the results of which are presented in \cite{NIM} and show that the on ground performance of the instrument is well understood and can be accurately reproduced using Monte Carlo simulations. Furthermore the performance of the instrument in-orbit and potential differences with those measured on ground were studied in detail. An example of one of the results in shown in figure \ref{bg_mod} which shows the amplitude of the 180 degree modulation measured by POLAR from the background as a function of its position in-orbit. This results show that although only a small amplitude is measured, which has the potential to induce a fake polarization signal, the amplitude of this modulation is very stable and not dependent on for example the albedo induced background which shows a strong correlation with magnetic latitude. This indicates that the background modulation can be corrected for with relative ease in the analysis phase. All the results of the in-orbit performance of POLAR are described in great detail in \cite{ZL} which shows that the in-orbit performance is now well understood. With the finalization of these instrument performance studies the focus can shift to the analysis of the polarization of the brightest GRBs detected by POLAR, the first catalog is expected to be published in early 2018 with more in depth papers focusing on individual GRBs following in subsequent months.

\section{Conclusions}

POLAR is a scintillator based instrument dedicated to measuring the polarization of the GRB prompt emission in the 50-500 keV energy range. The instrument was launched successfully as part of TG-2 on the 15th of September 2016 followed by 6 months of successful data taking during which a total of 55 GRBs were detected. The sample of measured GRBs contains approximately 10 GRBs for which a measurement with an MDP below $30\%$ can be achieved. For the strongest GRBs detected by POLAR the acquired statistics could also suffice to for example perform time dependent polarization studies. In April 2017 the instrument stopped data taking due to a problem with the high voltage power supply, attempts to recover this system are still ongoing. Despite the problems with the high voltage power supply the first half year of operation will allow to produce a first catalog of detailed polarization measurements which is expected in early 2018.

\end{document}